# Optical Band Engineering of Monolayer WSe$_2$ in a Scanning Tunneling Microscope

Xuehao Wu[1†], Thomas P. Darlington[1,2†], Madisen A. Holbrook[1,2], Emanuil S. Yanev[2], Luke N. Holtzman[2], Xiaodong Xu[3,4], James C. Hone[2], D. N. Basov[1*], P. James Schuck[2*], Abhay N. Pasupathy[1,5*]

[1] Department of Physics, Columbia University, New York, NY, 10027, USA
[2] Department of Mechanical Engineering, Columbia University, New York, NY, 10027, USA
[3] Department of Physics, University of Washington, Seattle, Washington, 98195, USA
[4] Department of Materials Science and Engineering, University of Washington, Seattle, Washington, 98195, USA
[5] Condensed Matter Physics and Materials Science Division, Brookhaven National Laboratory, Upton, USA

∗Corresponding authors. Email: db3056@columbia.edu, p.j.schuck@columbia.edu, apn2108@columbia.edu

†These authors contributed equally to this work.

*Intense electromagnetic fields can result in dramatic changes in the electronic properties of solids. These changes are commonly studied using optical probes of the modified electronic structure. Here we use optical-scanning tunneling microscopy (optical-STM) equipped with near-field continuous wave (CW) laser excitation to directly measure the electronic structure of light-dressed states in a monolayer transition metal dichalcogenide (TMD) semiconductor, WSe$_2$. We find that the effective tunneling gap and tunneling density of states are strongly influenced by the intensity of the electromagnetic field when the applied field is resonant with the bandgap of the semiconductor. Our findings indicate that CW laser excitation can be used to generate light-dressed electronic states of quantum materials when confined strongly to the nanoscale.*

Intense electromagnetic fields can be used to tune the electronic properties of solids, giving rise to new emergent electronic properties and functionalities[1,2]. Recent examples include the observation of the optical Stark effect[3,4], Floquet replicas of band structure[5-7], photo-induced magnetism[8] and superconductivity[9,10], and the tuning of the band gap of two-dimensional (2D) semiconductors under intense light fields[11]. Most of these experiments utilize pulsed laser fields where strong intensities can be achieved transiently. The electronic properties of the photoexcited state are also typically measured via optical methods due to the fast time scale of these measurements.

Near-field optical techniques using metallic tips are an alternative technique to achieve strong light intensities[12] while simultaneously achieving sub-diffraction optical resolution. A prominent example of this is scanning near-field optical microscopy (SNOM) which has undergone many refinements since the original work by Pohl[13]. In recent incarnations, the scanning probe of an AFM is used for apertureless SNOM, with the spatial resolution arising from the strong enhancement of the electric field near the sharp tip of the probe[14-16]. The illumination itself is in the far field, and geometric constraints of the tip and scanning apparatus typically limit the light field intensity at the tip. The use of plasmonic enhancement and nano-optical cavities[12,17-19] is an exciting development in this regard, but it remains challenging to study the nanoscale electronic response to intense electromagnetic fields in solids. Such measurements are crucial, however, for inducing and probing light-dressed states in solids that have disorder or nanoscale structure.

In this work, we use a modified version of a scanning tunneling microscope (STM) to probe the electronic properties of 2D semiconductors under strong optical excitation with scanning tunneling spectroscopy (STS). Our approach employs optical fiber coupled probes with plasmonic materials to deliver large photon fluences directly to the STM tunnel junction[15]. The technique also allows for collection of scattered and emitted light (Fig. 1a)[20]. We show in this work the ability of this probe to dramatically influence local electron tunneling in the valence band of monolayer $WSe_2$ (1L-$WSe_2$).

**Optical-STM evidence for high fields in a nanocavity**

In this experiment, we study nanobubbles of the 2D semiconductor 1L-$WSe_2$[21], which is a prototypical semiconducting TMD. Our sample is produced by direct exfoliation onto a template-stripped (TS) Au substrate[22,23] (more fabrication details available in the SI). The template stripping naturally forms regions of atomically flat crystalline terraces. Nanobubbles of $WSe_2$ are formed naturally during the exfoliation process (Fig. 1b) on these terraces. These nanobubbles avoid metal-induced quenching of photoluminescence (PL) and can form strong plasmon-exciton polariton states through interactions of excitons in the nanobubble and the nanocavity[24]. Fig. 1b, c show STM topographic images of a typical nanobubble, and of the $WSe_2$ lattice and Moiré pattern on gold. The performance of the STM itself is comparable to a traditional metal tip-based instrument.

Typically, optical coupling to the tunnel junction is achieved by focusing light via an external optic like a single lens or parabolic mirror. While this is a versatile way to couple light across the spectrum to the junction[25-27], it is limited by the geometry of the tip and scanning probe apparatus which makes it hard to achieve high numerical apertures. Our probe consists of a tapered single mode optical fiber (Fig. 1a) which allows us to deliver near-field laser excitation and realize in-situ collection. The tapered fiber is partially coated with gold to provide the current feedback signal for traditional STM/STS measurements and it also allows light delivery/collection through the uncoated portion of the tapered surface, with plasmonic guiding down to the apex of the tip[28]. When the Au-coated optical-STM probe approaches the surface with Au substrate, the probe-sample junction forms a plasmonic cavity and confines light to ultra-small mode volumes[29]. These highly confined interaction volumes establish the key feature for many cutting-edge nano-optical experiments, for instance enabling sub-nm-resolution Raman and STML measurements on molecules[30-35], as well as forming the backbone for surface-enhanced and tip-enhanced Raman spectroscopies (SERS and TERS, respectively). The achievable intensity enhancements within these zeptoliter mode volumes can be greater than $10^9$ [36], resulting in measured electric field strengths of ~1 – 20 MV/cm using only a HeNe 633 nm laser source[37]. Our apparatus allows us to collect light through the fiber, which can originate either from STM current-induced luminescence (STML) or from nano-photoluminescence (nano-PL). Fig. 1d shows STML and nano-PL spectra measured from the center of a nanobubble. The STML spectrum exhibits an emission peak at a slightly different energy than the nano-PL. The nano-PL spectrum under high laser field features a double peak, which may be attributed to the coexistence of exciton and trion. While the STML spectrum is dominated by a single emission peak of the trion due to the background electron population as discussed further below.

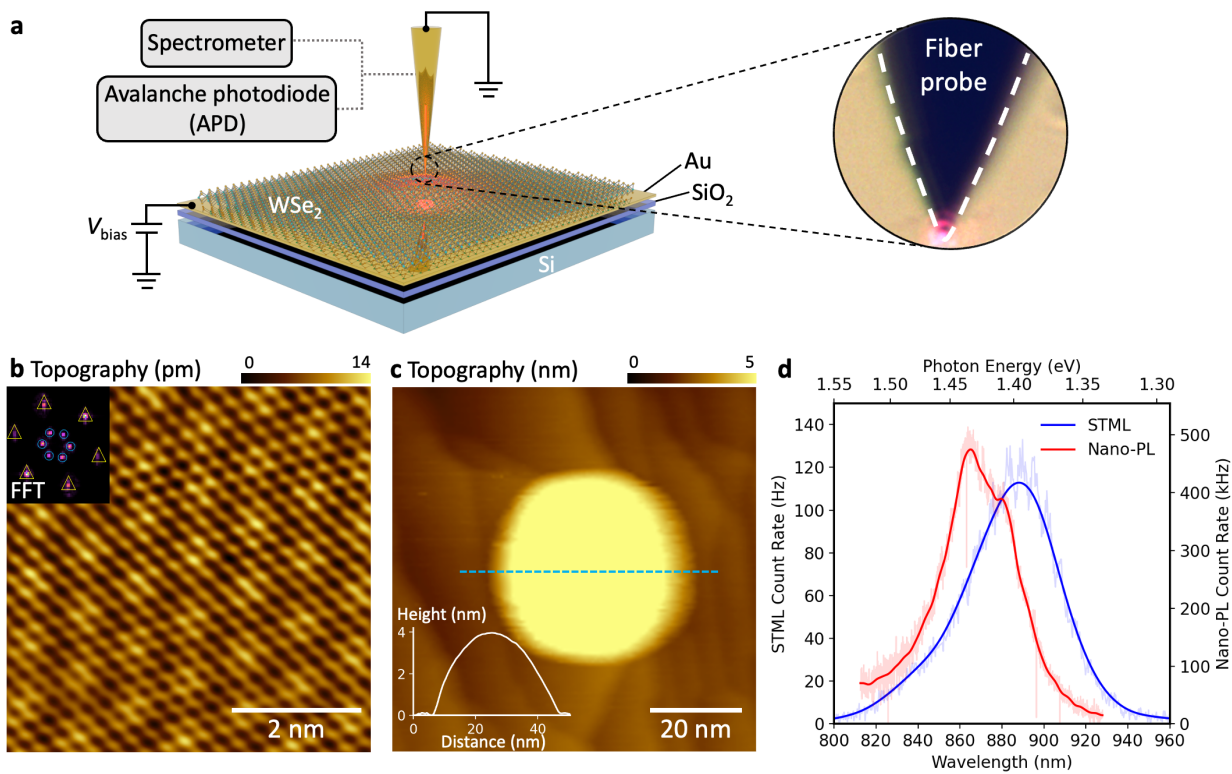

**Figure 1. Optical-STM measurement of light-dressed states in 1L-WSe$_2$ Nanobubbles:** (a) Schematic of our optical-STM showing the fiber probe with optical illumination, 1L-WSe$_2$ sample and external spectrometer/avalanche photodiode (APD). Inset: Optical image of the zoomed-in view of the probe-sample junction in operando. (b) Atomically resolved STM topography of 1L-WSe$_2$ on Au substrate. ($V_{bias}$ = 0.7 V, $I_{set}$ = 250 pA). Both the atomic structure of 1L-WSe$_2$ and the Moiré pattern arises from 1L-WSe$_2$/Au substrate are resolved. Inset: FFT of the topography with atomic peaks (yellow triangles) and Moiré peaks (blue circles) marked. (c) STM topography image of a typical 1L-WSe$_2$ nanobubble. Inset: linecut over the nanobubble as dashed blue line indicates. ($V_{bias}$ = 1 V, $I_{set}$ = 10 pA). (d) STML and Nano-PL spectra collected from a 1L-WSe$_2$ nanobubble under STM tunneling current excitation ($V_{bias}$ = -3 V, $I_{set}$ = 1 nA) and 633 nm CW laser excitation, respectively.

The intensity of the total STML observed from the nanobubble has a pronounced bias dependence, as illustrated in the simultaneously obtained STS and STML shown in Fig 2a. The STS without illumination shows the conduction band edge (CBE) and valence band edge (VBE) at around energies $V_{CBE}$ = -0.35 V and $V_{VBE}$ = -1.5 V, respectively, indicating electron doping of our 1L-WSe$_2$. The STML signal is only observed for V < $V_{VBE}$. The fact that the STML is not observed for V > $V_{CBE}$ can arise from the small thermal population of holes, fast non-radiative decay pathways or a combination of both effects. Conversely, for hole tunneling ($V_{bias}$ < $V_{CBE}$), the presence of STML indicates prominent exciton formation due to the background population of electrons which can recombine with the holes tunneled from the tip. The same bias dependence of STML has been reported in recent experiments on monolayer MoSe$_2$[38]. Comparison between the STML and STS spectra in Fig. 2a further confirms that $V_{bias}$ = -1.5 V is the K-point of the valence band of 1L-WSe$_2$, as has been previously discussed in STM measurements[39,40].

We next investigated the effect of 633 nm CW laser illumination on the STS, a wavelength chosen since it is close to the bandgap of pristine 1L-WSe$_2$. We performed STS measurements at the center of a 1L-WSe$_2$ nanobubble under moderate laser excitation powers (~1.8 µW CW laser excitation; see SI). Fig. 2b shows a large bias range STS with and without laser illumination. Surprisingly, a prominent enhancement of dI/dV in the valence band is observed under laser excitation. In contrast, the conduction band shows a negligible modification. We further explore this observation below.

In Fig. 2c, d, we compare the STM topography and spatial STML image collected from the same nanobubble. The STML image exhibits a donut-like shape, indicating that the STML from the edge of the nanobubble is stronger than the center (Fig. 2e). In contrast, the STM topography (Fig. 2f) and nano-PL image (Fig. 2g) collected from another nanobubble exhibit stronger nano-PL signal localized at the center of the nanobubble (the results are reproducible across nanobubbles). The nano-PL image linecut shows a significantly smaller diameter $d_{nano-PL} \sim 40$ nm compared to the diameter in topography $d_{topo} \sim 60$ nm (Fig. 2h). The difference between STML and nano-PL can be explained by the strain localization in TMD nanobubbles[41] which yields a smaller bandgap and induces higher electron density around the periphery. Such an electronic localization enhances the STML during the hole-tunneling process and suppresses the PL quantum yield when it is pumped optically. Here the sharp sub-10 nm resolution of nano-PL image clearly demonstrates the nanoscopic nature of the scanning plasmonic cavity and associated enhancement formed at the tip-sample junction.

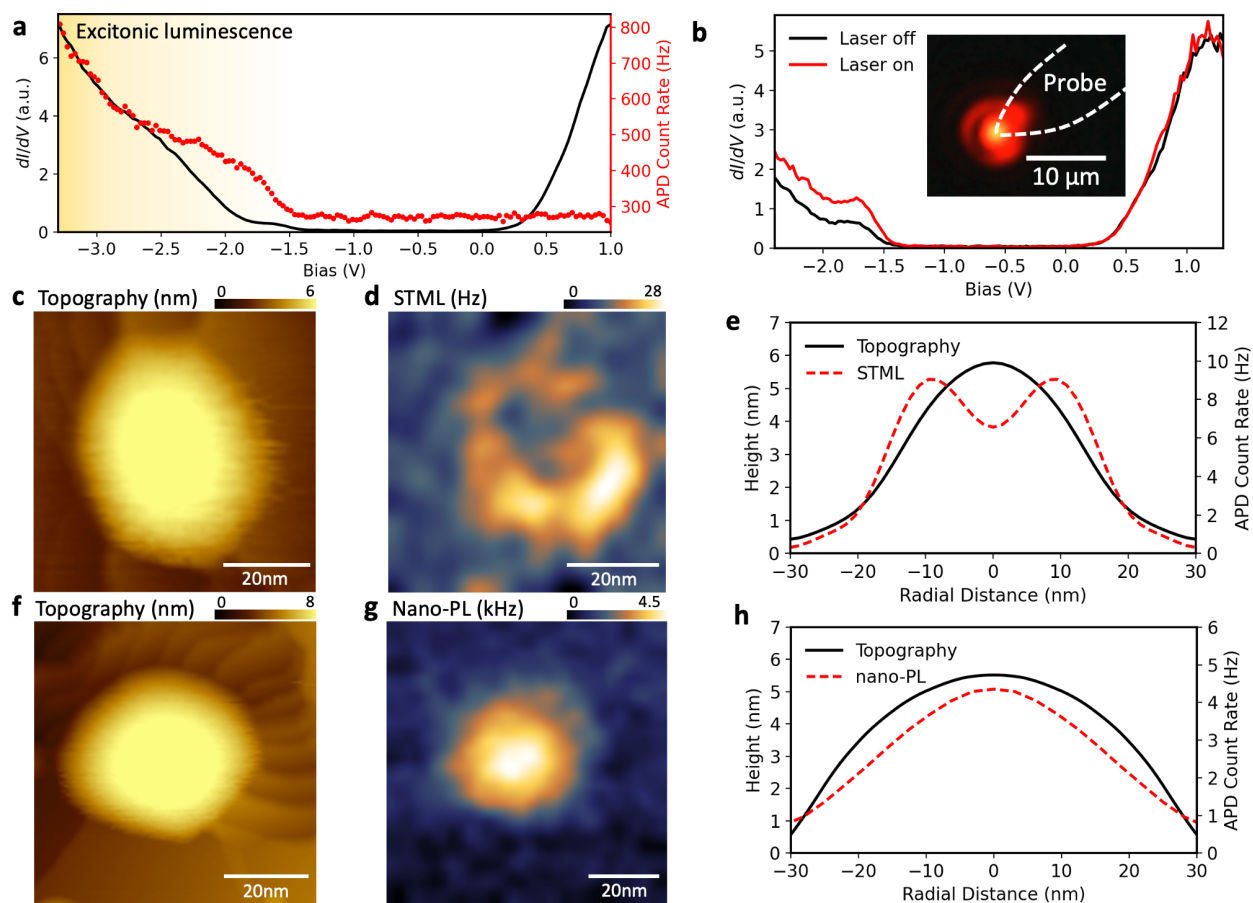

**Figure 2. Excitonic luminescence under tunneling current excitation (STML) and near-field CW laser excitation (nano-PL) and STS with laser turned off and on in a 1L-WSe$_2$ nanobubble:** (a) STS (black) and STML (red) spectra collected at center of a 1L-WSe$_2$ nanobubble simultaneously. (b) STS spectra measured at center of a 1L-WSe$_2$ nanobubble with 633 nm CW laser excitation turned off (black) and on (red). Inset: optical image of the optical-STM probe with laser on. ($V_{bias}$ = 1.3 V, $I_{set}$ = 100 pA). (c) STM topography image of a 1L-WSe$_2$ nanobubble. ($V_{bias}$ = 1.4 V, $I_{set}$ = 10 pA). (d) STML spatial imaging of the same nanobubble in (c) ($V_{bias}$ = -2.7 V, $I_{set}$ = 150 pA). (e) Comparisons of topography and STML radially averaged signals in (c)(d). (f) STM topography image of a 1L-WSe$_2$ nanobubble. ($V_{bias}$ = 1.4 V, $I_{set}$ = 10 pA). (g) Nano-PL imaging of the same nanobubble in (f) with 100 µW 633 nm CW laser illumination. (h) Comparisons of topography and nano-PL radially averaged signals in (f)(g).

**Light-dressed states indicated by STS**

To better understand the results of Fig. 2b, we performed STS measurements as a function of laser fluence on a single nanobubble (Fig. 3a). As the fluence increases, a strong increase in dI/dV is observed in the valence band, whereas the conduction band remains unchanged. The strongest dI/dV enhancement is observed at $V_{bias} \sim -1.5$ V, where the onset of the STML was observed (Fig. 2a). This suggests that the feature is dominated by K-point states. Moreover, a clear valence band peak (VBP) feature develops in the STS around this bias at powers 1 - 20 µW. At higher power, the VBP appears to merge into deeper valence band features.

In Figs. 3b, c, we further investigate the behavior of the VBP. The tunneling conductance at the VBP increases dramatically with laser power up to 10 µW, over a factor of 10 compared with the STS without laser excitation (black curve in s. 3b). We also evaluate the modification of the bandgap of the 1L-WSe$_2$ nanobubble while tuning the CW laser power (Fig. 3d). The bandgap is defined by the energy difference between the conduction band edge (CBE) and valence band edge (VBE) as determined by polynomial fits (see SI for detailed discussion). Under low power excitation (< 4 µW), the observed bandgap increases monotonically by ~20 meV. Under stronger excitation up to 100 µW, the trend reverses, with the bandgap decreasing by ~70 meV. The striking fluence dependence of the STS is our major scientific finding.

Such strong modification of the tunneling spectrum under optical illumination is unprecedented in STM studies of two-dimensional semiconductors. However, in the context of optical studies, power-dependent quantum properties of WSe$_2$ are well established[42], particularly the properties of many types of excitons that the material can host leading to density-dependent phases[43]. As such, several possible mechanisms can participate in the phenomena we observe, which we discuss in more detail below.

At low laser power (< 4 µW), one of the prominent effects we expect is laser-induced Pauli blocking[44] – the laser creates a steady state population of holes and electrons at the valence and conduction band edges respectively, and these states are then unavailable for STM tunneling. This mechanism naturally leads to a larger effective band gap as measured by STS. At larger laser powers, the steady state population of excitons begins to change the tunneling gap due to the enhanced screening. This bandgap renormalization is a well-documented phenomenon in optical measurements[45,46].

While the changes in the size of the measured tunneling gap can be reasonably attributed to Pauli blocking and bandgap renormalization, these effects cannot explain the strong enhancement in the dI/dV signal at the valence band edge. We first address the possibility that this enhancement is a systematic artifact, driven by extrinsic phenomena such as thermal expansion of the tunneling junction. When the laser power is changed, each of our measurements is performed after allowing the junction to equilibrate for several hours. We see no difference in atomic scale STM topography of nanobubbles at the indicated laser fluences. Further, we do not observe a significant change in the slope of the STS at the CBE, which would be a signature of strong temperature dependence. All of these indicate that thermal effects are not at play. A related possibility is a photocurrent artifact. Photocurrents can change the junction normalization condition, which in turn can change the magnitude of dI/dV signals. In this regard, it is important to note that our junction normalization condition (V=0.7 V and I =30 pA in Fig. 3) keeps the conduction band dI/dV constant. The observed changes to the valence band cannot therefore arise from a change of the overall STS normalization.

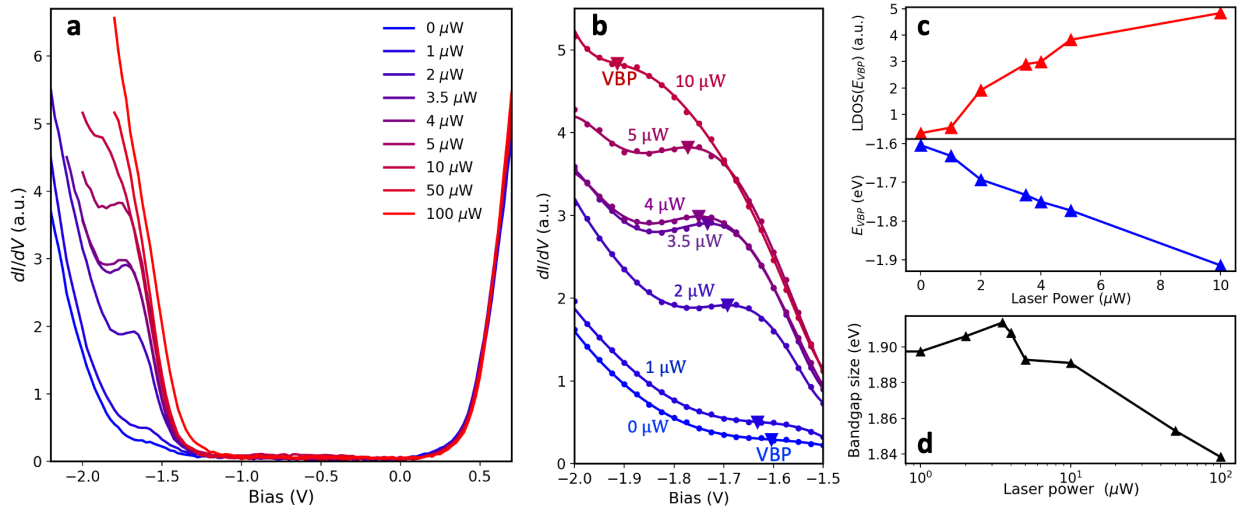

**Figure 3. Laser-induced light-dressed STS in a WSe$_2$ nanobubble:** (a) Laser power-dependent STS spectra (V$_{bias}$=0.7 V, I$_{set}$=30 pA) collected at the center of a 1L-WSe$_2$ nanobubble. (b) 6$^{th}$ order polynomial fitting of STS spectra of valence band with fitted valence band peaks (VBP) marked as triangles. (c) Top panel: local density of states (LDOS) at valence band peak vs. laser power. Bottom panel: valence band peak energy vs. laser power. (d) Bandgap size of the 1L-WSe$_2$ nanobubble in (a) vs. laser power.

One important question that arises is whether the observed enhancement of valence band dI/dV is dependent on the wavelength of light used. To investigate this, we repeated our experiments at laser wavelengths of 750 nm, 790 nm, and 850 nm (well below the band gap). Almost no modification of the LDOS is observed at the valence band with those off-resonance laser energies, in strong contrast to the case of 633 nm (1.96 eV) excitation we employ in Fig. 3, suggesting a resonance effect is at play (full data and discussion in SI).

**Numerical simulation of STS**

In Fig. 4, we explore potential mechanisms for the modification of the STS that would be expected under strong optical pumping. They can be sorted into two types of effects: electric field effects and band effects as indicated in Fig. 4. We simulate the modification of LDOS by following a model[47] that employs a perturbative treatment called "Bardeen approach" for each situation. Under electric field effects, we consider changes in the effective height or thickness of the tunnel barrier as illustrated in Fig. 4a and b. We also consider the possibility of laser-field-induced bias rectification (Fig. 4c), which has already been observed in ultrafast STM measurements[18]. While the exact magnitude of these effects is unknown since they depend on the details of the tip shape, we simulate tunnel barrier parameters that encompass all reasonable parameters. None of these situations can yield a modification of the LDOS similar to what we observed. (Detailed discussion in SI)

We next consider the bandgap renormalization effect (Fig. 4g), which can ultimately lead to Mott transitions at very high exciton densities[45,46]. In general, this effect can cause changes in the chemical potential of the semiconductor. Our experiments show that the shape and energy position of the conduction-band derived dI/dV does not change with laser power. Bandgap renormalization effects therefore can only result in a shift of the valence band edge as shown in Fig. 4i. The clear changes in spectral shape observed in our measurements cannot be explained by this phenomenon.

Another mechanism we explore is the effect of a change in band dispersion of the valence band at the K-point caused by the electromagnetic field. For a 2D semiconductor, the LDOS is proportional to the effective mass. The observed enhancement of LDOS can indicate band flattening at the 1L-WSe$_2$ K-point upon higher laser illumination. Previous optical pump-probe experiment[3] on monolayer WS$_2$ has demonstrated that the hybridization between light-induced Floquet-Bloch bands and equilibrium Bloch bands can result in energy repulsion between them, which is known as the optical Stark effect[48,49]. Such an energy repulsion is proportional to the light intensity and can yield a flatter band edge that hosts a larger effective mass. Since the excitation in our experiment is CW, and to date nearly all studies on optical Stark effect require high pulse intensities, this possibility may seem implausible. However, the high field enhancement afforded by the plasmonic nanocavity can increase field strengths to above 1 MV/cm[50], with values regularly in the 1-20 MV/cm range for previous plasmonic-gap-mode tip-enhanced studies[36]. Similar cavity Floquet engineering of 1L-WSe$_2$ at low fluences has been reported recently[4]. Moreover, previous optical-STM measurements have demonstrated Floquet-like shifting of field emission resonances (FER) by a photon quantum in similar plasmonic nanocavities under laser illumination[17], showing that the field intensities can be sufficient to realize optical engineering through the optical Stark effect. This is schematically shown in Fig. 4h. Since the energy of our laser is roughly on resonance with the bandgap of the nanobubbles, as revealed by the STS in Fig. 2, the light-dressed state of the conduction band $|c - hv>$ will have significant hybridization with the valance band state at the K-point. If the interaction is strong enough, this hybridization can effectively flatten out the top of the valence band. Recent tr-ARPES observations on black phosphorus have shown that strong pumping of intense IR pulse at resonance can lead to Floquet band engineering, leading to a larger effective mass[5] which is similar to our observation. Fig. 4j illustrates the expected STS for three different values of the effective mass of the valence band. Additionally, for large changes in the effective mass, a clear peak emerges on the VBE due to the change in band dispersion near the band edge. Of the five situations we considered in Fig. 4, changes in the effective mass most closely resemble our experimental results in Fig. 3. In this situation, the absence of modification of the CBE can be explained by the asymmetric scattering of photogenerated carriers in the valence and conduction bands, as high scattering is well known to destroy the hybridization to light-dressed states[6]. For 1L-WSe$_2$, this can be facilitated by the nearby Λ valley in the conduction band (Fig. 4h), which provides a much closer scattering state, compared with the well isolated K-valley in the valence band.

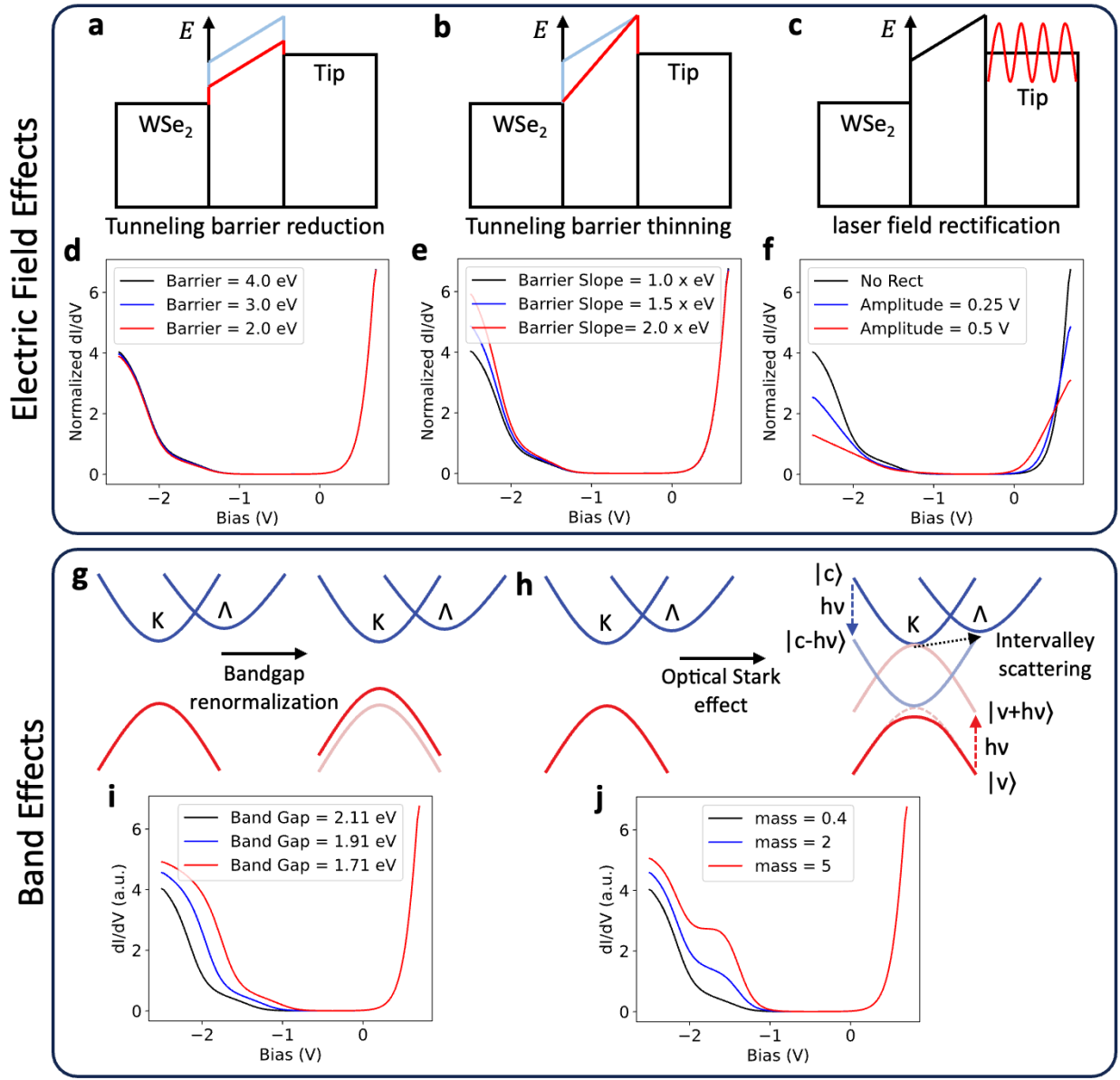

**Figure 4. Potential mechanisms for STS modification in 1L-WSe$_2$ that arise from electric field effects and band effects:** (a)(b)(c) Diagrams of three potential electric field effects: tunneling barrier reduction, tunneling barrier thinning and laser field rectification, respectively. (d)(e)(f) Simulated LDOS modification correspond to (a)(b)(c) based on perturbative STS model[47]. (g)(h) Diagrams of two potential band effects: bandgap renormalization and optical Stark effect, respectively. (i)(j) Simulated LDOS modification correspond to (g)(h).

**Conclusion**

In conclusion, using a newly developed optical-STM we demonstrate the electroluminescence and nano-PL nature of 1L-WSe$_2$ nanobubbles and investigate effects of light-dressing on the 2D electronic states. We achieved STML with tunneling current perturbation and sub-10 nm resolution of nano-PL imaging under laser illumination with the optical-STM technique. We further investigated the electronic properties of the light-dressed states by studying STS spectra as a function of laser intensity. The observed behavior of the LDOS in the valence band is suggestive of band hybridization with light-dressed bands through optical Stark effect, opening an experimental route to understanding and exploiting the fascinating regime of CW optical band engineering. More generally, these capabilities of optical-STM provide us opportunities to explore nano-optical response and light-dressed phases of quantum matters with STM in sub-nanometer spatial resolution, unlocking applications for on-demand quantum engineering with light.

**Acknowledgements**

This work is supported by Programmable Quantum Materials, an Energy Frontier Research Center funded by the U.S. Department of Energy (DOE), Office of Science, Basic Energy Sciences (BES), under award DE-SC0019443. Support for development of the optical fiber tips is provided by the Air Force Office of Scientific Research under grant FA9550-16-1-0601 (A.N.P.)

**Author contributions**

Optical-STM measurements: XW, TPD
Sample fabrication: ESY, TPD
Data analysis: XW, TPD
WSe$_2$ synthesis: LNH
Manuscript writing with input from all authors: XW, TPD, DNB, PJS, ANP

**Methods**

**Sample preparation:** Template stripped gold (TS Au) substrates were prepared via a "cold-welding" process to avoid uses of epoxies. Two wafers of Au were prepared: a thinner layer of Au (~20 nm) on top of a 5 nm Ti adhesion layer and another with a thicker (~100 nm) pure Au layer. Diced chips (1 cm x 1 cm) were placed together in a vice using 2-3 layers of Tek paper to achieve uniform loading over the chip area. Chips were removed after ~ 1hr with a razor blade, which left a uniform Au film with a rms roughness of the underlying wafer, the typical rms is smaller than 0.5 nm. WSe$_2$ monolayers were exfoliated directly by using Scotch tape onto TS Au chips immediately after stripping to minimize surface contamination. Presence of nanobubbles in monolayer WSe$_2$ is verified by hyperspectral photoluminescence mapping using the LabRAM HR Evolution Raman microscope from Horiba Scientific.

**STM and STS measurements with laser excitation:** STM and STS measurements were performed in an ultra-high vacuum chamber. dI/dV spectroscopy measurements were performed with a lock-in amplifier. A continuous wave HeNe laser source with a wavelength of 633 nm was coupled to our fiber probe to excite the sample. We placed a set of neutral density filters outside the STM chamber to control the laser excitation power for the power-dependent

STS measurements. Each set of STM and STS data was conducted several hours after we adjusted the laser power to eliminate the influence of thermal drift.

**STML and nano-PL measurements:** STML (STM tunneling current excitation) and nano-PL (laser excitation) signal are collected through the fiber probe and sent to an avalanche photodiode (APD) and a spectrometer for analysis. STML and nano-PL imaging measurements were conducted by APD while scanning for STM topography. STML and nano-PL optical spectra were measured by the spectrometer while the tip is parked typically for two minutes.